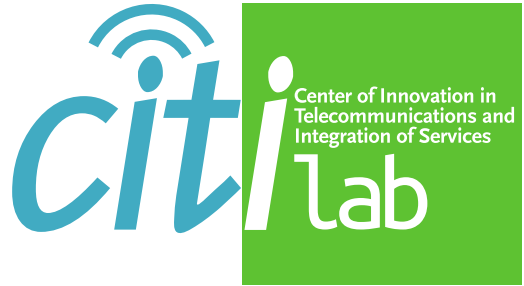

# Technical Report

Center of Inovation in Telecommunications and Integration of services

# Radio-Enabled Low Power IoT Devices for TinyML Applications

**Antoine Bonneau** [1, 2]   **Fabien Mieyeville** [2]   **Frédéric Le Mouël** [1]

**Régis Rousseau** [1]

[1] Univ Lyon, INSA Lyon, Inria, CITI, EA3720, 69621 Villeurbanne, France
*{antoine.bonneau, frederic.le-mouel, regis.rousseau}@insa-lyon.fr*

[2] Univ Lyon, Université Claude Bernard Lyon 1, INSA Lyon, Ecole Centrale de Lyon, CNRS, Ampère, UMR5005, 69622 Villeurbanne, France
*fabien.mieyeville@univ-lyon1.fr*

December 06, 2023

**ABSTRACT**   The proliferation of Internet of Things (IoT) devices has intensified the demand for energy-efficient solutions supporting on-device and distributed learning applications. This research presents a circumscribed comparative analysis of radio-enabled ultra-low power IoT devices, specifically focusing on their suitability for computation-heavy use cases. Our analysis centers on middle-end IoT devices that serve as a vital interface between the Electronics and Machine Learning communities. The evaluation encompasses a diverse range of IoT hardware equipped with integrated radios. We established functional datasheet-based criteria completed with accessibility and community-wise criteria to study and offer valuable insights into each selected node's performance trade-offs, strengths, and weaknesses. This study provides crucial guidance for TinyML practitioners seeking to make informed device selections for their applications.

# I   INTRODUCTION

The convergence of Edge Computing (EC) [1] and the Internet of Things [2] has catalyzed a paradigm shift in connected devices. The traditional centralized computing paradigm, commonly associated with cloud systems, has gradually expanded towards terminal devices. By distributing computational tasks across many compact devices strategically positioned at the network's periphery, EC minimizes data transfers to distant cloud servers, resulting in significant cost savings and improved efficiency. Its distributed nature means that the failure of individual nodes can be gracefully managed with minimal disruption to the overall system. Within this context, we have selected noteworthy communicating and computing boards, offering a wide array of available protocols. We considered hardware with a configurable reduction in energy consumption while maintaining high performance for Machine Learning (ML) use cases [3]. This is achieved by providing significant persistent and working memory capabilities. The advancements in the microelectronics industry underpinning these hardware improvements benefit all practitioners wishing to design long-lasting environment-embedded ML solutions [4]. This report aligns with existing work on the subject, categorizing IoT devices into three distinct tiers we recall and update in Section II, each tailored to specific application needs and resource constraints. Our focus is directed towards ready-to-use development boards for quick prototyping of applications with demanding profiles (like TinyML) and a need to communicate at lower cost [5]. Therefore, we have listed interesting characteristics for this type of application in Section III. Subsequently, we enumerate a certain number of devices meeting these criteria, detailing the differences and properties by which they distinguish themselves in Section IV. Finally, in Section V, we provide avenues for reflection on hardware and software improvements to explore to reach the latest levels of edge intelligence [6].

# II   TERMINOLOGIES AND CONCEPTS

## II.A   Constrained IoT Node Classification

The Internet of Things (IoT) field encompasses a diverse range of devices tailored to meet specific application needs. Devices within this domain are often categorized based on hardware specifications, connectivity options, processing power, and data processing sophistication. Literature usually divides IoT devices into three categories, each designed to address distinct levels of complexity and functionality in IoT applications: *Low-end*, *Middle-end*, and *High-end* from the simplest to the most complex tier. The most relevant recent studies on the subject are those by Ojo et al. [7], which cover the various IoT device classes, exposing the devices used by researchers, and by Wisniewski et al. [8], which focuses on the power consumption of edge computing devices. However, these two studies have different classifications. Indeed, Ojo et al. use an IETF standardized classification [9] as a reference for Low-end devices, defining High-end devices as having enough power to run traditional operating systems like Linux and place Middle-end IoT devices filling in between. Wisniewski et al. rely on the classification proposed by Neto et al. [10], which focuses on the abstraction levels at which systems can handle information extraction and fusion. This classification considers a broader range of devices, so it is not easy to be precise about low-performance appliances. The approaches are respectively oriented from the perspective of the electrical engineers and the ML specialists, leaving disparity in definitions. We propose reconciling the classifications by relying, in addition to these works, on the revision of RFC 7228 [11], although this is a work in progress.

## 1. Low-end IoT devices

Low-end IoT (LeIoT) devices encompass a category characterized by simplicity and resource constraints. These devices feature primary temperature, humidity, and motion sensors, making them suitable for straightforward applications like environmental monitoring and home automation. Their computing power, memory, and storage capacities are minimal. Instead, they typically use low-level firmware or dedicated Wireless Sensor Networks (WSN) operating systems for programming. At the same time, these devices are cost-effective and well-suited for lightweight data processing. LeIoT devices generally operate at a low abstraction level, often using programming models closer to the hardware. Table 1 shows the three different classes of the low-tier devices of IoT based on their memory budget and resulting software constraints. We have reconstructed it from the work of Ojo et al., and extracted it from the recommendations of the IETF working group. Even if Class 0 nodes are not sufficiently equipped to embed network stacks up to the application layer, they can still be designed to utilize at least the physical and data link layers.

| Class Name | Memory Size | | RTOS Support | Communication protocols |
|---|---|---|---|---|
| | Data e.g. RAM | Code e.g. Flash | | |
| Class 0 C0 | $\ll$ 10 KiB | $\ll$ 100 KiB | Devices do not support RTOS | No protocol stack embedded, gateways for communications |
| Class 1 C1 | $\simeq$ 10 KiB | $\simeq$ 100 KiB | Custom RTOS implementation | Lightweight protocols such as CoAP. They do not need gateways to communicate |
| Class 2 C2 | $\simeq$ 50 KiB | $\simeq$ 250 KiB | Minimal RTOS operation | General communication protocols are supported |

**Table 1.** Classes of low-end IoT devices (KiB = 1024 bytes)

## 2. Middle-end IoT devices

Middle-end IoT (MeIoT) devices offer a balanced blend of performance, features, and cost-effectiveness. They boast advanced sensors, diverse connectivity options, and more capable processors. Compared to low-end devices, MeIoT devices offer greater capabilities with multiple communication technologies, higher clock speeds, and increased RAM. These devices can store, collect, and process data, often with monocore microcontrollers or application cores. They provide comprehensive functionality beyond basic statistics processing. Table 2 summarizes the rest of the RFC 7228bis [11] classes in the microcontroller (M) group. These classes are less constrained compared to the LeIoT ones. Devices in this class can easily run lightweight operating systems but still face energy supply limitations and can hardly meet the extreme requirements for a complete data processing pipeline usually required by advanced AI solutions until full training.

| Class Name | Memory Size | | OS Support | ML Algoithms |
|---|---|---|---|---|
| | Data e.g. RAM | Code e.g. Flash | | |
| Class 3 C3 | ≃ 100 KiB | ≃ 500..1000 KiB | RTOS are fully supported | ANN inference |
| Class 4 C4 | ≃ 300..1000 KiB | ≃ 1000..2000 KiB | RTOS are encouraged | ANN training |

**Table 2.** Classes of middle-end IoT devices (KiB = 1024 bytes)

## 3. High-end IoT (HeIoT) devices

High-end IoT (HeIoT) devices are specialized devices with ample computing resources, including robust processing units, substantial RAM, and substantial storage capacity, often complemented by a Graphical Processing Unit. These devices can execute traditional operating systems and perform resource-intensive tasks like running heavy ML algorithms. They feature comprehensive on-board connectivity options, encompassing FastEthernet/GigaEthernet interfaces, WiFi/BT chipsets, and HDMI outputs. They frequently incorporate camera interfaces to support multimedia applications. Table 3 gathers the highest classes as defined by Bormann et al. [11] and covers the remaining spectrum of possible memory capacities. These classes no longer represent constrained IoT nodes at all but serve to expand the range of devices within the digital ecosystem of modern networks.

| Class Name | Memory Size | | Device Example |
|---|---|---|---|
| | Data e.g. RAM | Code e.g. Flash | |
| Class 10 C10 | ≃ 16..128 MiB | ≃ 4..16 MiB | OpenWRT routers |
| Class 15 C15 | ≃ 0.5..1 GiB | ≃ 16..64 MiB | Raspberry PI |
| Class 16 C16 | ≃ 1..4 GiB | *(lots)* | Smartphones |
| Class 17 C17 | ≃ 4..32 GiB | *(lots)* | Laptops |
| Class 19 C19 | *(lots)* | *(lots)* | Servers |

**Table 3.** Classes of high-end IoT devices (KiB = 1024 bytes)

## II.B Low Power Considerations

The specifications of an embedded electronic board to be considered low power (LP) are typically challenging to quantify in numerical terms. It is more of an approach to circuit design [12] that involves using components to minimize current leakage. Measuring values vary considerably depending on the circuit's size and the number of components employed. There are no fixed values for embedded development boards used in IoT applications; instead, an average can be obtained by aligning experimental conditions within each class. LP and ultra-low-power (ULP) boards use energy-efficient components such as low-leakage transistors and, more specifically, low-power CMOS techology [13], frugal microcontrollers, and dedicated sensors to minimize power consumption.

These boards employ advanced power management techniques like dynamic voltage scaling, clock gating, and power gating to optimize power consumption based on the workload. LP research aims to design communicative devices that incorporate a specific computational capacity yet function autonomously without reliance on electrical grids or conventional batteries.

| | | | | ✓ On | ✗ Off | |
|---|---|---|---|---|---|---|
| **Mode** | **Core Status** | **Storage Status** | **RAM Status** | **Peripherals Status** | **Power Consumption** | **w.r.t to Idle Mode** |
| Active with Transmission (TX) | ✓ | ✓ | ✓ | ✓ | ≤ 1 W | 1000 % |
| Active with Reception (RX) | ✓ | ✓ | ✓ | ✓ | ≤ 500 mW | 500 % |
| Active | ✓ | ✓ | ✓ | radio is sleeping | ≤ 200 mW | 200 % |
| Idle | slower clock rate | slower clock rate | ✓ | radio is sleeping | ≤ 100 mW | 100 % |
| Light Sleep | ✗ | ✗ | limited subset | own clock based only | ≤ 10 mW | 10 % |
| Deep Sleep | ✗ | ✗ | ✗ | ✗ | ≤ 100 μW | 1 % |
| Power Off | ✗ | ✗ | ✗ | ✗ | ≤ 100 nW | 0.05 % |

**Table 4.** Order of magnitude of consumption for common MeIoT operating modes

Table 4 represents the orders of magnitude we obtain for the MeIoT on the selected manufacturers and is provided for informational purposes only. It is likely to decrease with the advancements in microelectronics research. The transmission and reception correspond to the states of the integrated radio when it is switched on to exchange packets using the supported protocols. These are the most power-hungry states that one would seek to minimize to the absolute minimum. The *Active* mode corresponds to the MCU mode used to compute. *Idle* mode is an alternate mode when the core is awake but waiting for peripherals to recover from queried instructions. *Light Sleep* corresponds to a minimal board status, possibly gathering environmental information without computing, and *Deep Sleep*, where MCU is off, is the default mode to conserve energy while limiting wake-up time. The *Power Off* mode corresponds to the current leakages when the device is fully shut down, but the power source is still wired. All those modes are constructor-dependents and are not standardized yet.

## III CRITICAL CONSIDERATIONS EVALUATING IOT HARDWARE

Several considerations come into play when evaluating hardware for Internet of Things (IoT) applications. Quantitative and Qualitative aspects, respectively Section III.A and III.B, can be highlighted to help different project holders decide. While the trend leans towards using ARM architectures, we aim to spotlight other chip designers to broaden the choices for future generations and open a high-potential alternative path against the giant ARM with its licensed instruction set architecture (ISA). It would be unfortunate if a monopoly were to impede progress in the sector just as competitors begin to emerge. Among these, RISC-V is a rapidly accelerating player [14], offering a free instruction set and fostering initiatives in open hardware. It is worth noting, however, that despite the apparent advantages of RISC-V, there currently needs to be more support compared to the dominant ARM architecture in the embedded systems domain. Architectural choices impact the ability of IoT devices to meet the requirements of the state-of-the-art TinyML algorithms in their respective domains. Considering the requirements exhibited in these studies surveyed by Abadade et al. [3], and the fact that the nodes should also have memory space for network stacks and operating systems, the targeted devices are roughly those in the Middle-end IoT category.

### III.A QUANTIFIED PERFORMANCE CRITERIA

Performance criteria, when quantified, offer a systematic approach to assessing the capabilities and features of electronic devices and systems. These quantifiable metrics are appropriate tools for informed decision-making, helping select devices that align most effectively with the unique demands of a particular project or application.

#### 1. PROCESSING AND MEMORY CAPABABILITIES

The choice of the microcontroller unit (MCU) architecture is a critical aspect of LP design. The selection may include 16 to 64-bit MCUs. Each architecture has its trade-offs regarding computational power, code size, and power efficiency. An important feature is the adaptive scaling clock speed range. Lower clock speeds generally result in reduced power consumption. The clock speed is typically kept as low as possible for LP boards while still meeting the application's performance requirements. The amount of RAM, ROM, and Flash available on the board impacts the system's power consumption and performance. A board with limited RAM and ROM might be more power-efficient, but it could also limit the complexity of applications it can handle. The right balance between the different memory types is necessary to achieve an LP operation while supporting the desired functionalities.

#### 2. PERIPHERALS AND WIRED COMMUNICATIONS

Intelligent nodes require diverse peripherals to engage with their environment effectively. These systems rely on a versatile set of interfaces, encompassing everything from General-Purpose Input/Output (GPIO) pins to plug additional modules to Analog-to-Digital Converters (ADC) for precise data acquisition. The current trend in sensor technology involves integrating the instrumentation and conversion components directly within the sensor itself rather than delegating this task to the microcontroller. The GPIOs are also used for wired interactions with other devices through interfaces like the Serial Peripheral Interface (SPI) and Inter-Integrated Circuit (I2C). These interfaces depend on the microcontroller chosen by the card manufacturer and are therefore fixed, but they enable communication with additional hardware such as external sensors, peripherals, and communication modules. However, it is imperative to consider that each interface carries its unique power consumption profile. Although most commercially available MCUs can accommodate a wide range of these interfaces, they are often selectively disabled when not in use. This is a prudent strategy for conserving resources on custom-designed boards.

### 3. Radio Tranceivers and Wireless communications

The radio transceiver module enables wireless communication in LP systems. The choice of the radio transceiver should consider the specific communication protocols supported and their power consumption, as battery-operated devices must conserve energy during wireless data transmission. Many communication protocols are available, each tailored to different needs and use cases. The most commonly cited ones include 6LoWPAN, Wi-Fi, Zigbee, Bluetooth, Bluetooth Low Energy (BLE), and manufacturer custom 2.4 GHz protocols. For instance, Wi-Fi is known for its high data rates and robustness but can be power-hungry. In contrast, BLE is designed with LP consumption. Conversely, Zigbee compromises data rate and power efficiency while allowing mesh deployment. These protocols differ in data transmission capabilities, range, data rate, and power consumption. We target boards with integrated radio to facilitate newcomers' adoption.

### 4. Electrical characteristics

Battery-operated devices rely on specific voltage ranges and current consumption for optimal performance. Efficient LP boards are designed to operate effectively at lower voltages, thus extending battery life. Precise voltage control is essential to prevent damage while minimizing current consumption is vital for prolonged battery life and reduced heat generation. The manufacturer sets the supply voltage, which can be regulated. However, the current draw varies depending on the running application's activated peripherals and available power-saving modes.

### 5. Unit Price

The board's cost is a substantial factor in the decision-making process for research and commercial endeavours. LP boards must find an equilibrium among performance, features, and cost-efficiency. The board's price plays a role in assessing whether the solution is economically viable for large-scale production, research endeavours and do-it-yourself (DIY) projects democratizing embedded AI. However, it may be more advantageous to consider designing a custom board at this juncture.

### 6. Product dimensions

In applications where space is a precious commodity, the physical dimensions of the board assume relative importance. Achieving a compact form factor is often imperative for LP boards utilized in wearable devices, IoT sensors, or small-scale embedded systems. These dimensions are commonly expressed as length by width in millimetres (mm), while the height is commonly around 5mm.

### III.B Qualitative evaluation parameters

We adopt a holistic perspective beyond technical specifications and numerical metrics in our approach to evaluating IoT device selection criteria. We focus on delving into qualitative dimensions, which provide insights into the human and critical practical factors crucial in selecting IoT devices for various applications and projects. These parameters are closely or remotely related to the device's helpful information and easy access.

## 1. Energy-aware Features

Energy-efficient features on IoT boards are purposefully crafted to optimize power usage. These encompass elements like current measurement pins within Microcontroller Units (MCUs), a range of supply options such as battery sockets, and practical solutions like reversible solders or jumpers that allow the deactivation of power-draining unused components. These thoughtful features empower IoT devices to maximize their energy resources, rendering them suitable for various applications and environments while significantly prolonging their battery life.

## 2. Technological Maturity

The maturity of an IoT board reflects its development status, testing history, and real-world application. Boards that have demonstrated success in diverse applications and earned popularity among hobbyists and industries are typically favoured. Academic research and literature can further reinforce the effectiveness of specific boards, especially in LP applications. Additionally, assessing the board's release date and monitoring manufacturer updates provides insight into its ongoing relevance and the commitment to continuous improvement and support. However, the recent release of the card should be nuanced as there are microcontrollers dating back to the early 2000s that are still in use today without substantial improvements. For instance, the ATMEGA 1281, released in 2005, equips today's Libelium waspmotes [15]. We chose to represent this criterion by emphasizing the developer communities for which the considered card might offer a comparative advantage in terms of invested time. The profile of contributing agents and project support reflect this balance.

## 3. Community Activeness

Community activity within embedded software development encompasses several vital aspects that denote its vibrancy and effectiveness. The support structure within the community, including forum activity and size, articles referencing the hardware or software, and the rate of resolution for feature requests within the codebase, are crucial indicators. Additionally, the richness of examples manufacturers and users provide, such as tutorials, boilerplates, and test codes consistently updated and compatible with the latest hardware and software, speaks volumes about the community's engagement. Another vital aspect lies in community libraries—evaluating their relevance, volume, availability for specific hardware, ease of use, maintainability, and adaptability across various frameworks. These libraries play a pivotal role in streamlining development processes and fostering cross-framework projects, contributing significantly to the collaborative and innovative spirit of the embedded software community.

## 4. Documentation Quality

High-quality documentation for boards and MCUs is an invaluable asset for IoT projects, simplifying development by offering clear information on pin configurations, interfaces, and functionalities, thus streamlining design and coding. Moreover, comprehensive documentation often includes examples, code libraries, and practical usage guidelines, which expedite development and support the learning curve for newcomers to IoT. The quality of documentation significantly influences the success of IoT projects, providing developers with essential tools and guidance to bring their ideas to fruition.

Adequate IoT board documentation typically encompasses two to three technical documents focusing on the MCU, board specifications, and the development environment.

### 5. Development Tools accessibility

Simplicity and user-friendliness are top priorities, especially for IoT developers with varying expertise levels. IoT boards should offer straightforward setup procedures and intuitive interfaces. Additionally, compatibility with popular development environments like Arduino [16] or PlatformIO [17] projects or the availability of extensions for commonly used Integrated Development Environments (IDEs) enhances the overall development experience for IoT enthusiasts. However, in practice, manufacturers often create proprietary software development tools with varying levels of quality.

## IV    OFF-THE-SHELF IOT HARDWARE SOLUTIONS

In the field of embedded computing, it is imperative to recognize that devices emanating from the same generation of microcontrollers generally exhibit comparable energy consumption characteristics, as suggested by the research conducted by Khouloud et al. [18]. Nevertheless, it is incumbent upon us to acknowledge that divergences in the design of the electronic circuit board can induce disparities in energy consumption, even when identical microcontrollers are employed. These discrepancies may be attributed to component layout, interconnection pathways, or ancillary components. We are considering boards capable of running multiple wireless protocols in order to facilitate experimentation. Table 5 provides a qualitative evaluation summary of the studied boards, reflecting the criteria outlined in Section III.B. Conversely, Table 6 presents a quantitative analysis summary based on the criteria specified in Section III.A for the same set of boards.

✓ Good    ✗ Bad    ∼ Partial    — Missing

| IoT Board | Energy-Aware Features | Technological Maturity | | | Documentation Quality | | | Community Activeness | | |
|---|---|---|---|---|---|---|---|---|---|---|
| | | Hobby | Academy | Industry | MCU | Board | IDE | Libraries | Support | Samples |
| ESP32 DK M1 | Solder bridges | ✓ | ✓ | ✓ | ✓ | ✓ | ✓ | ✓ | ✓ | ✓ |
| ESP32 C6 DK C1 | MCU current measurement connector | ✓ | ✗ | ✗ | ✓ | ∼ | ✓ | ∼ | ∼ | ✓ |
| STM32 WB55RG | MCU current measurement connector, Peripherals jumpers, Battery socket | ✗ | ✗ | ✓ | ✓ | ∼ | ∼ | ✗ | ✗ | — |
| Arduino Nano 33 BLE | Solder bridges | ✓ | ✓ | ✗ | ✓ | ∼ | ✓ | ✓ | ✓ | ∼ |
| nRF 7002 DK | MCU current measurement connector | ✗ | ✓ | ✓ | ✓ | ✓ | ✗ | ∼ | ✓ | ✓ |
| LaunchPad CC1352P7-4 | Peripherals jumpers | ✗ | ✓ | ✗ | ✓ | ✗ | ✓ | ✗ | ∼ | — |

**Table 5.** Radio-enabled IoT devices qualitative criteria overview

| IoT Board | Application Core | | Network Core | | Regulated / Unregulated / Sleep (min) / Transmission (max) | | | | |
|---|---|---|---|---|---|---|---|---|---|
| | Processing Unit | Clock Speed (MHz) | Memory (kB) | On-board Storage (kB) | Electrical Characteristics | Wireless Connectivity | Low-level Peripherals | Release & Price* | Size L × W (mm) |
| ESP32 DK M1 | Xtensa LX6 dual-core 32-bit | 80 to 240 | 520 (RAM) 448 (ROM) | 4096 | R 3.3 ± 0.3 V U 5..12 V S 5 µA T 379 mA | WiFi 5 Bluetooth 4.2 BLE | 28 GPIO pins, 12 bits ADC, UART, SPI, SDIO, I²C, PWM, I²S | Dec. 2020 8 € | 51×26 |
| ESP32 C6 DK C1 | RISC-V 32 bit | 80 to 160 | 400 (RAM) 384 (ROM) | 4096 | R 3.3 ± 0.3 V U 5..12 V S 7 µA T 354 mA | WiFi 5 BLE (mesh) 802.15.4 | 22 GPIO pins 12 bits ADC 3 SPI, 2 UART, I²C, LPI²C, I²S, PCNT, RMT, DMA, SDIO, USB, 2 TWAI, PARLIO, ETM | Jan. 2023 9 € | 60×26 |
| STM32 WB55RG | Arm Cortex M4 32-bit | 64 | 256 (RAM) | 1024 | R 5 ± 0.25 V U 7..12 V – | Bluetooth 5.2 802.15.4-2011 | 68 GPIO pins, 12 bits ADC, DMA, USB, (Q)SPI, SAI, (LP) UART, I²C | Sep. 2018 33 € | 70×57 |
| Arduino Nano 33 BLE | Arm Cortex M4F 32-bit | 64 | 256 (RAM) | 1024 | R 3.3 V U 12..21 V – | Bluetooth 5 BLE 802.15.4 NFC-A | 23 GPIO pins, 12 bits ADC, EasyDMA, (Q)SPI, TWI, PDM, QDEC USB, I²S, I²C | Apr. 2021 27 € | 45×18 |
| nRF 7002 DK | Arm dual Cortex M33 32 bit | A 160 N 64 | A 1024 N 256 | 4096 | R 3.3 ± 0.3 V U 5..12 V S 15 µA T 260 mA | WiFi 6 BLE 802.15.4 NFC | 31 GPIO pins, DPPI, EGU, NFCT, NVMC, (Q)SPI, I²C, USB, UART, EasyDMA | Jan. 2023 52 € | 137×64 |
| LaunchPad CC1352P7-4 | Arm Cortex M4F 32-bit | 48 | 152 (RAM) 256 (ROM) | 704 | 2.8 ± 1.0 V | Bluetooth 5 BLE 802.15.4 Wi-SUN 6LoWPAN … | 26 GPIO pins, 12 bits ADC, 2 UART, I²S, I²C, 2 SSI | Oct. 2020 62 € | 120×59 |

\* As of December 06, 2023

**Table 6.** Radio-enabled IoT devices quantitative criteria overview

## IV.A Espressif Family

Espressif Systems[1] is a Chinese fabless semiconductor company known for its innovative microcontroller and connectivity solutions. Their flagship products have garnered significant attention and acclaim within the hobbyist and research communities. Espressif Systems fosters a supportive ecosystem for developers to create and experiment with cutting-edge IoT and wireless communication applications. The manufacturer's software suites *ESP-IDF* [19] and *ESP-Arduino* [20] are highly intuitive, with frequent updates and rapid support for new boards. Users have access to numerous open-source code samples that demonstrate the capabilities of their technology. The company focuses on SoCs embedding a microcontroller and a radio combined on a single chip and provides a solid foundation for a wide range of applications. Their products are undeniably the most cost-effective, but their radios consume significantly more energy.

---

[1] https://www.espressif.com

### 1. ESP32 Development Kit

We adopt a holistic perspective beyond technical specifications and numerical metrics in our approach to evaluating IoT device selection criteria. We focus on delving into qualitative dimensions, which provide insights into the human and practical factors that play a pivotal role in selecting IoT devices for various applications and projects. ESP32 DK is an entry-level development board [21]. This Espressif product stands as a testament to a decade-long journey. Its robust design and versatile functionality have solidified its reputation as a go-to choice in the IoT community. This record of exadaptability and reliability, honed over the years, makes the ESP32 a remarkable reference among hobbyists and industrial users in the modern era. It is powered by the outsider MCU Xtensa LX6 dual-core 32-bit, first announced in 2004. Its SoC [22] has a reasonably large amount of RAM and supports Wi-Fi and Bluetooth without needing an external radio.

### 2. ESP32 C6 Development Kit

ESP32-C6 DK is Espressif's first Wi-Fi 6 SoC integrating 2.4 GHz Wi-Fi 6, Bluetooth 5, BLE and the 802.15.4 protocol. ESP32-C6 achieves an emerging industry-leading RF performance with reliable security features and multiple memory resources for IoT products. Its SoC [23] consists of a high-performance (HP) 32-bit RISC-V processor, which can be clocked up to 160 MHz. It has a 320KB ROM and a 512KB SRAM and works with external flash. It directly improves the ESP32-C3 [24] architecture, sharing many common features. The SoC documentation is available, but specific documentation for the board still needs to be included. Indeed, the recent release of this model is currently being embraced by hobbyist, industrial, and academic communities. The C6 version is currently unstable due to its recent release, unlike C3, which was released in 2022. This instability is reflected in external projects that typically target Espressif products.

## IV.B  STM32 Wireless Family

STMicroelectronics[2] (STM) is a renowned global semiconductor company at the forefront of innovation in microcontroller technology, offering an extensive portfolio of embedded solutions. Their STM32 series, in particular, stands out as a versatile and powerful microcontroller family catering to the needs of industrial applications. The STM32 series features a range of microcontrollers equipped with high-performance LP Arm MCUs. It is important to note that certain aspects of their documentation need more in-depth examples. Moreover, the Eclipse-based software suite *CubeMX/CubeIDE* [25], intended to ease the onboarding of new users, tends to become cumbersome and limiting when developing custom applications. This proprietary environment lacks support and code snippets that accelerate prototyping. Its closed-source model generally hinders developers seeking to design an advanced system. There is an open-source initiative [26] aiming to provide Rust interfaces for ST's cards.

### 1. Nucleo WB55RG

The Nucleo WB55RG [27] is a development board featuring a 32-bit Arm Cortex-M4 based processor [28] released in 2017. This board includes various jumper options for configuring the microcontroller and measuring power consumption. It offers support for a CR2032 battery, enhancing its power supply versatility. This board has 68 GPIO pins, allowing interaction with many external peripherals. The node embeds a ULP device embedding a frugal radio compliant with the BLE and 802.15.4 protocols.

---

[2] https://www.st.com

## IV.C Nordic Semiconductor Family

Nordic Semiconductor[3] stands out as a leader in wireless communication and LP microcontroller solutions. Its expertise in power-efficient wireless technologies, integrated software stacks, and a rich ecosystem of development tools make it an appealing choice, particularly for IoT and wearable device development. The size and price of these development boards make them difficult to use outside prototyping. The latest boards focus on acquiring and manipulating audio processing but can also be used within a TinyML context. A lot of their SoCs have been used to develop formidable IoT boards.

### 1. Arduino Nano 33 BLE

Arduino[4] offers various products focusing on education but is used by a broad range of hobbyists, students, and academicians. One of Arduino's most significant strengths is its accessibility. It is widely recognized for its beginner-friendly approach, characterized by simplicity in both hardware and software aspects. The platform also benefits from an extensive ecosystem of compatible sensors, shields, and add-on modules, streamlining the rapid prototyping of diverse projects. Nevertheless, its processing power and wireless integration limitations can be significant obstacles. Few boards have LP focus and good radio integration. There may be better choices for developing commercial or mass-market products than Arduino. In addition, their IDE is highly praised by the maker community for its simplicity and the extensive number of supported modules and boards; many boards from other manufacturers can also be programmed using it.

The Arduino Nano 33 BLE [29] is a compact and versatile development board ideal for IoT and wireless communication projects. It features a powerful 32-bit ARM Cortex-M4 processor and onboard BLE capabilities manufactured by Nordic [30], making it suitable for a wide range of applications, from sensor monitoring to wearable devices. Other boards from their Nano series are intriguing, and some products from the MKR lineup, while not explicitly designed to minimize power consumption, have their own merits. The XIAO BLE nRF52840 Sense board [31] is an alternative that also uses the eponymous SoC released in 2018 and promises Zigbee integration.

### 2. nRF 7002 Development Kit

The nRF7002 DK [32], [33] is the first device in Nordic's portfolio of unique Wi-Fi products that will combine seamlessly with Nordic's existing ULP technologies. Nordic brings their decades of LP wireless IoT and silicon design expertise to Wi-Fi. The sixth version of the protocol benefits IoT applications, including further efficiency gains supporting long-life, battery-powered Wi-Fi operation. The DK combines the nRF5340 DK [34], [35] capabilities with a Wi-Fi 6 companion Integrated Circuit. The MCU was first released in 2020. This predecessor also features a battery socket facilitating wireless use case evaluation. The two boards have pins for sharp monitoring of SoC power consumption. The board is powered by two 32-bit Cortex M33 MCUs offering parallelization for application and networking routines with standard and mutual memory units.

---

[3] https://www.nordicsemi.com
[4] https://www.arduino.cc/

## IV.D Texas Instrument Family

Texas Instruments (TI)[5], is a prominent semiconductor company renowned for its comprehensive range of products. TI strongly emphasizes energy-efficient solutions, making their MCUs ideal for battery-powered and IoT applications. TI's development kits are a bit light on memory, but some boards have been used extensively in the LeIoT tier. On the other hand, they have very few devkits supporting their most interesting radio-enabled microcontrollers. Their *Code Composer Studio* [36] and SimpleLink LP SDKs [37] are old and closed-source but functional.

### 1. LaunchPad CC1352P7-4

The CC1352P7-4 LaunchPad [38] is a dual-band LPradio kit that leverages the CC1352P7 chip released in 2020. It features a 32-bit ARM Cortex-M4F processor and few memory capacities compared to other MeIot Boards. This development kit supports wireless protocols, including BLE, Zigbee, and Thread. It is a modern alternative to their highly popular MSP430 [39], even if we deplore the absence of FRAM, an invaluable feature of its predecessor. It is a pity that the MCU documentation is precise, but that of the LanchPad is missing. This device has a dedicated section on TI's support forum [40], which holds the majority of available online support. However, it is sparsely populated, containing approximately 150 topics.

## V PERSPECTIVES TOWARDS IN-SITU OPERATION

In the ever-evolving technological landscape, a pressing concern arises at the intersection of energy management and the growing need for real-time data acquisition in IoT devices. Addressing this issue requires exploring new ways to power our devices, as outlined in Section V.A, adapting communication, synchronization, and scheduling for a system prone to frequent failures, as presented in Section V.B, and adopting a new method of computation to rapidly and efficiently process information extracted from the environment, introduced in Section V.C.

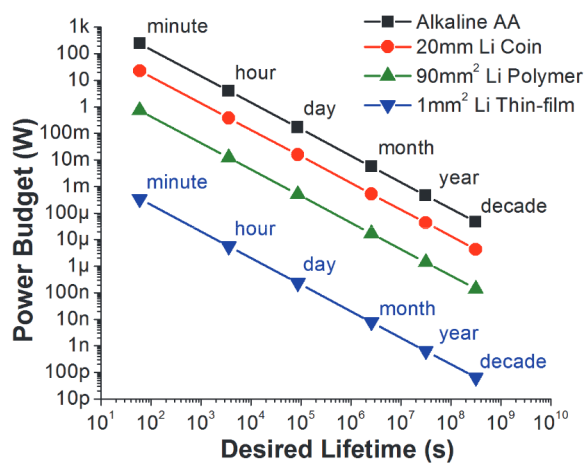

**Figure 1.** Average power draw constraint as a function of lifetime and battery size [41]

---

[5] https://www.ti.com

## V.A ENERGY HARVESTING

Energy conservation and sustainability [42] is a fundamental concern in the design of IoT devices, especially those destined for remote or hard-to-access environments. These devices' small-sized batteries typically have capacity limitations thus rapid depletion, as illustrated in Figure 1 that lead to a restricted operational lifespan. Consequently, energy recovery, often called Energy Harvesting (EH), assumes critical significance. EH [43] involves retrieving and utilizing ambient energy sources, including solar energy, mechanical vibrations, ambient heat, or kinetic energy, to power IoT devices and supplement the operational lifespan of these devices. Various energy harvesting strategies can be tailored to specific environmental and device characteristics. Paramount energy recovery methods encompass using solar panels to capture solar energy or applying piezoelectric converters to convert mechanical vibrations into electrical energy.

| Harvesting Method | Working principle | Power density | Characteristics | | |
|---|---|---|---|---|---|
| | | | Ambient | Controllable | Predictable |
| Solar Energy (outdoors) | Photovoltaic | 15-100 mW/cm³ – bright day<br>0.15-100 µW/cm³ – cloudy day | ✓ | ✗ | ✓ |
| Solar Energy (indoors) | | 6 µW/cm³ | ✗ | – | ✓ |
| Mechanical Energy, Vibration, Motion | Microgenerator | 800 µW/cm³ (human activity – Hz)<br>4 µW/cm3 (mechanism – kHz) | ✗ | ✓ | ✗ |
| | Piezoelectric | 200-330 µW/cm³ | ✗ | ✓ | ✗ |
| | Electrostatic | 184 µW/cm³ | ✗ | ✓ | – |
| Temperature Variations | Pyroelectric | 10 µW/cm³<br>(Temperature rate 8.5°C) | ✗ | ✗ | ✗ |
| Ambient RF | Rectenna | 0.08 nW - 0.1 µW/cm³<br>(GSM 900/1800 MHz)<br>0.01 µW/cm³ (Wi-Fi) | ✓ | ✗ | ✓ |
| EM Waves | Coils | 125 µW/cm³ (2.65 Hz) | ✓ | ✗ | ✓ |
| Hydro energy, Wind, Airflow | Turbine, Triboelectric, Generator | 65.2 µW/cm³ (5 m/s)<br>400 µW/cm³ (TEG & TENG) | ✓ | ✗ | ✗ |
| Acoustic Noise | Diaphragms, PZT | 0.003 µW/cm³ (75 dB)<br>0.96 µW/cm³ (100 dB) | ✓ | ✗ | ✗ |

**Table 7.** Performance of harvestable ambient energy sources with power density from [43]

Table 7 offers a comprehensive view of the efficiencies achievable through energy harvesting solutions. Because it is the easiest to harness, the predominant energy source is the recovery of photovoltaic cells when deployed outdoors in favourable weather conditions. Nonetheless, these ideal conditions are impractical for the year-round operation of our devices, including during winter, nighttime, and on days with poor weather, where energy recovery is notably diminished. The selected embedded systems operate on the magnitude order of a few square centimetres, significantly limiting the amount of energy we can expect to harvest through energy harvesting. At this scale, energy harvesting typically does not generate enough power to sustain the nodes continuously, even if designed to be highly energy-efficient, as highlighted by the work of Kausar et al. [44]. Therefore, it becomes imperative to integrate an energy storage capacity within the system to smooth and extend energy availability. This energy storage capability will be crucial in mitigating energy production fluctuations while also considering the need to effectively manage power interruptions [45], ensuring continuous and reliable operation of our embedded systems.

EH kits are readily available for microcontrollers and are provided by specialized manufacturers such as Seedstudio [46] at competitive prices. These kits typically encompass a compact photovoltaic panel, a 500mAh LiPo battery, and a dedicated board featuring an integrated circuit for monitoring battery charge and discharge. Additionally, these components are available for individual purchase, offering flexibility in tailoring solutions to specific project requirements. Customization options include selecting the appropriate solar panel size to suit project needs or exploring alternative energy harvesting modules. Furthermore, the battery can be modified and accompanied by supplementary cells or supercapacitors for enhanced capacity.

The pivotal element within these kits is the charger built around the buck converter and LDO components. They are designed to adeptly manage multiple storage components and potentially diverse energy sources. Some chargers incorporate advanced functionalities like Maximum Power Point Tracking (MPPT) [47], optimizing power extraction by aligning with the energy source's unique characteristics. Affordable options offered by Adafruit [48] (housing a TI chip) and more expensive boards, such as the e-peas AEM10941 evaluation board [49] are compatible with the microcontroller cards outlined in Section IV. These solutions boast ease of implementation, eliminating the need for soldering and functioning almost as plug-and-play systems. Additionally, monitoring the battery's charge level via I2C communication protocol is feasible, depending on the chosen kit.

Regrettably, robust commercial cards specifically tailored for this equipment are lacking. It is essential to exercise caution when choosing energy storage technology for deployments that require prolonged operation without human intervention. Parameters such as temperature operating ranges and charge cycles demand careful consideration from the initial stages of the design process.

## V.B Embedded Operating Systems

An embedded Operating System or IoT OS [2], [50] is a specialized software that manages the resources and provides essential functionalities for embedded devices. Unlike general-purpose OSs, IoT OSs are tailored for specific hardware configurations and constrained environments, optimizing performance and resource utilization. They facilitate communication between hardware and higher-level software, handling tasks like memory management, task scheduling, and device drivers and providing a framework for applications to run efficiently on embedded systems. Adopting an IoT OS is particularly relevant when integrating intricate libraries or protocol stacks, as third-party software often operates through multiple threads. One could still manage by using the APIs specific to the manufacturers, but it would require a lot more effort to build a complex system and be limited to cards compatible with those APIs. Embedded OSs are diverse and tailored to suit the intricate demands of embedded systems across industries. Real-time OS (RTOS) ensures precise timing, critical for applications like industrial automation or medical devices. Linux-based embedded OS provides versatility, supporting a wide range of hardware and software for complex applications but with a significant overhead. Proprietary OSes offer optimized performance for specific hardware but with limited flexibility. IoT OSs allow working with a low footprint at an additional level of abstraction.

| IoT OS | Platforms | RAM (kB) | ROM (kB) | Codebase (MLOC) | Programming Language | Scheduling | Simulation / Emulation | Last Release |
|---|---|---|---|---|---|---|---|---|
| Apache MyNewt | ARM (M0/3/4/23/33) MIPS32, Microchip PIC32, RISC-V | $\gtrsim 1$ | $\gtrsim 10$ | ~ 1.2 | C, C++ | Preemptive Priority based | – | 1.11.0 Sept. 2023 |
| Apache Nuttx | ARM, AVR8, MIPS, Renesas, RISC-V, Xtensa, ZiLOG | – | – | ~ 4.3 | C, C++ | Preemptive Priority based | QEMU | 12.2.1 Jul. 2023 |
| Zephyr OS | ARM, x86, ARC, RISC-V, Nios II, Xtensa, SPARC | – | ~ 2 to ~ 8 | ~ 3 | C, C++ | Preemptive, Priority based, Cooperative | QEMU native | 3.5.0 Oct. 2023 |
| FreeRTOS | ARM, AVR(32), ColdFire, Xtensa, HCS12, RISC-V, IA-32, Infineon XMC4000, MicroBlaze, MSP430, PIC(32), Renesas H8/S, RX100/200/600/700, 8052, TriCore, EFM32 | ~ 1 | ~ 5 to ~ 10 | ~ 7.2 | C, C++, Go (Rust Wrappers) | Ticless, Preemptive, Priority based, Cooperative | QEMU (Cortex M3) | 10.6.2 Nov. 2023 |
| RIOT OS | ARM, MSP430, MIPS, AVR, x86, RISC-V | ~ 1.5 | ~ 5 | ~ 3 | C, C++ (Rust Wrappers) | Tickless, Preemptive, Priority based | QEMU, Rednode | 2023.10 Aug. 2023 |

**Table 8.** IoT OSs Overview

Table 8 gives an overview of the Embedded OS landscape corresponding to our research criteria by displaying the characteristics retrieved from the documentation of each of them. We have also approximated the number of lines of code by retrieving files from their open-access core code base. This metric is highly uncertain since many files that are not directly linked to the functionality base can distort the evaluation. However, it gives a good approximation of the effort the developers put into the project and its inertia. The considered projects all support the microcontrollers we selected, even though minor adjustments may be required to support board specificities fully. We listed only non-proprietary and open-source frameworks, ensuring users can study, modify, and distribute the code. Open-source licensing promotes collaboration, transparency, and community-driven development.

### 1. Apache MyNewt and Nuttx

Apache Mynewt [51] is a notable modular OS tailored for IoT devices operating under power, memory, and storage constraints. With a small kernel size and a rich feature set, it offers simplicity and ease of use, enabling developers to prototype, deploy, and manage 32-bit microcontroller-based IoT devices effortlessly. It stands out for NimBLE, its open-source BLE stack reused through other OSs.

Apache Nuttx [52] is designed to support a rich, multi-threaded development environment for deeply embedded processors by offering implementations of standard POSIX OS interfaces. NuttX is a highly scalable RTOS, supporting processors ranging from tiny 8-bit to moderate 64-bit devices, achieving scalability through strategies like using many compact source files, linking from static libraries, high configurability, and employing weak symbols when available.

### 2. FreeRTOS

FreeRTOS framework [53], a real-time operating system kernel for embedded devices, is a crucial foundation for IoT applications. Developed initially by Richard Barry, it was designed for simplic-ity and portability. FreeRTOS is primarily written in C with architecture-specific scheduler routines in assembly language. Amazon currently supports this project but retains its open-source model. It supports POSIX threads, software timers, and a tick-less mode for LP IoT applications.

### 3. Zephyr RTOS

The Zephyr OS [54] is a project initiated by the Linux Foundation in 2016 as a scalable and modular RTOS in which multiple semiconductor industrial companies participate. Zephyr offers a small footprint, preemptive multitasking, and extensive device driver support. It boasts security enhancements like memory protection and isolation and a rich set of connectivity protocols, making it versatile for diverse embedded applications. All listed boards have a corresponding base implementation requiring only the adaptation of the latest version, except for the TI's development kit.

### 4. RIOT OS

RIOT [55] established in 2008, is an open-source, community-driven RTOS designed for IoT. It focuses on low-power devices, featuring a small footprint and efficient memory usage. RIOT supports numerous architectures, enabling seamless connectivity with IPv6, 6LoWPAN, and various wireless protocols. With preemptive scheduling, it emphasizes energy efficiency and offers a modular structure for easy extensibility, making it suitable for resource-constrained IoT deployments. RIOTs offer a uniform API across diverse hardware platforms (8 to 32-bit). It can run directly as a process on Linux or FreeBSD systems.

## V.C Intermittent Computing

Within embedded systems, maintaining a certain level of service quality while coping with interruptions is a central challenge. Even though these devices spend most of their time in a sleep state to conserve energy, optimizing the activation, computation, and data transmission phases is imperative. A thorough architectural consideration is necessary to preserve essential time and information during interruptions. Although specialized operating systems such as Sytare, designed by Gautier et al. [56], geared towards non-volatile memory optimizations, exist for managing intermittency, their specificity makes them less suitable for generalization and easy deployment on diverse hardware architectures. Research is underway to integrate energy management into existing operating systems, such as CARTOS by Karimi et al. [57], which added APIs to FreeRTOS. However, the authors still assume the availability of NVRAM, whereas manufacturers often directly solder the fastest read interfaces (Quad SPI, Octo SPI, etc.) to slower Flash chips, making it challenging to integrate these solutions into plug-and-play prototypes.

Consequently, one option is to handle intermittency at the application level [58], providing a more flexible and adaptable solution to address the challenges of interruptions. ULP coprocessors in the latest generation of microcontrollers add a dimension to intermittence by enabling operational multithreading and a fallback to two extremely low-power tasks before entering a sleep state to preserve memory, ultimately powering off in the worst-case scenario. Simultaneously, the Intermittent Computing field of research is currently being explored to integrate ML algorithms. These works [59], [60] proposing task-based solutions are a promising direction in which practitioners who take on the challenge of operating with transient power supplies will have to dig in.

## VI CONCLUSION

Integrating Machine Learning into IoT devices underscores a rapidly growing field, with active involvement from various research teams. Selecting the proper hardware for edge computing is a pivotal aspect of this journey. In this report, we have updated one classification of IoT nodes and offered insights into a carefully curated collection of boards suitable for kickstarting projects at the outermost edges of computing. We strongly advocate for pursuing multiple avenues to achieve sustainable ambient intelligence. We observe the absence of hardware in MeIoT that meets the constraints of memory and computation while providing integrated energy harvesting solutions and efficient non-volatile memory. These solutions would further accelerate prototyping for wireless TinyML applications. Manufacturers must allow for configuring the power supply's energy mix at the software level for optimal energy consumption and regulatory compliance. In this direction, it might be interesting to delve into embedded OS hacking to add this functionality and build an API that facilitates intelligent checkpointing to adapt to regular power failures.

## REFERENCES

[1] X. Kong, Y. Wu, H. Wang, and F. Xia, "Edge Computing for Internet of Everything: A Survey", *IEEE Internet of Things Journal*, vol. 9, no. 23, pp. 23472–23485, Dec. 2022, doi: 10.1109/JIOT.2022.3200431.

[2] S. Bansal and D. Kumar, "IoT Ecosystem: A Survey on Devices, Gateways, Operating Systems, Middleware and Communication", *International Journal of Wireless Information Networks*, vol. 27, no. 3, pp. 340–364, Sep. 2020, doi: 10.1007/s10776-020-00483-7.

[3] Y. Abadade, A. Temouden, H. Bamoumen, N. Benamar, Y. Chtouki, and A. S. Hafid, "A Comprehensive Survey on TinyML", *IEEE Access*, vol. 11, pp. 96892–96922, 2023, doi: 10.1109/ACCESS.2023.3294111.

[4] S. S. Saha, S. S. Sandha, and M. Srivastava, "Machine Learning for Microcontroller-Class Hardware: A Review", *IEEE Sensors Journal*, vol. 22, no. 22, pp. 21362–21390, Nov. 2022, doi: 10.1109/JSEN.2022.3210773.

[5] A. Bonneau, F. Le Mouël, and F. Mieyeville, "Addressing limitations of TinyML approaches for AI-enabled Ambient Intelligence", in *Workshop on Simplification, Compression, Efficiency and Frugality for Artificial intelligence (SCEFA), in conjunction with the European Conference on Machine Learning and Principles and Practice of Knowledge Discovery in Databases (ECML PKDD)*, Sep. 2023. Accessed: Nov. 28, 2023. Available: https://hal.science/hal-04250026

[6] P. P. Ray, "A review on TinyML: State-of-the-art and prospects", *Journal of King Saud University - Computer and Information Sciences*, vol. 34, no. 4, pp. 1595–1623, Apr. 2022, doi: 10.1016/j.jksuci.2021.11.019.

[7] M. O. Ojo, S. Giordano, G. Procissi, and I. N. Seitanidis, "A Review of Low-End, Middle-End, and High-End Iot Devices", *IEEE Access*, vol. 6, pp. 70528–70554, 2018, doi: 10.1109/ACCESS.2018.2879615.

[8] L. Martin Wisniewski, J.-M. Bec, G. Boguszewski, and A. Gamatié, "Hardware Solutions for Low-Power Smart Edge Computing", *Journal of Low Power Electronics and Applications*, vol. 12, no. 4, p. 61, Nov. 2022, doi: 10.3390/jlpea12040061.

[9] C. Bormann, M. Ersue, and A. Keränen, "Terminology for Constrained-Node Networks", May 2014. doi: 10.17487/RFC7228.

[10] A. Rocha Neto *et al.*, "Classifying Smart IoT Devices for Running Machine Learning Algorithms", Jul. 2018, doi: 10.5753/semish.2018.3429.

[11] C. Bormann, M. Ersue, A. Keränen, and C. Gomez, "Terminology for Constrained-Node Networks", Jun. 2022. Accessed: Nov. 28, 2023. Available: https://datatracker.ietf.org/doc/draft-bormann-lwig-7228bis

[12] B. Ivey, "Low-Power Design Guide", 2011. Accessed: Nov. 28, 2023. Available: https://ww1.microchip.com/downloads/en/AppNotes/01416a.pdf

[13] A. Chandrakasan, S. Sheng, and R. Brodersen, "Low-power CMOS digital design", *IEEE Journal of Solid-State Circuits*, vol. 27, no. 4, pp. 473–484, Apr. 1992, doi: 10.1109/4.126534.

Bonneau et al.: Radio-Enabled Low Power IoT Devices for TinyML Applications 19Bonneau et al.: Radio-Enabled Low Power IoT Devices for TinyML Applications 19

[14] R.-V. Community, "RISC-V Sees Significant Growth and Technical Progress in 2022 with Billions of RISC-V Cores in Market". Accessed: Nov. 28, 2023. Available: https://riscv.org/announcements/2022/12/risc-v-sees-significant-growth-and-technical-progress-in-2022-with-billions-of-risc-v-cores-in-market/

[15] "Waspmote Technical Guide - Architecture and system". Accessed: Nov. 28, 2023. Available: https://development.libelium.com/waspmote-technical-guide/architecture-and-system

[16] "Arduino Documentation". Accessed: Nov. 28, 2023. Available: https://docs.arduino.cc/

[17] "PlatformIO: A new generation toolset for embedded C/C++ development". Accessed: Nov. 28, 2023. Available: https://docs.platformio.org/

[18] A. Khouloud and F. Mieyeville, "Développement de carte d'instrumentation pour l'analyse de puissance d'IoT pour du calcul distribué", 2020.

[19] "Espressif IoT Development Framework". Accessed: Nov. 28, 2023. Available: https://github.com/espressif/esp-idf

[20] "Arduino core for the ESP32". Accessed: Nov. 28, 2023. Available: https://github.com/espressif/arduino-esp32

[21] "ESP32-DevKitM-1 Programming Guide". Accessed: Nov. 28, 2023. Available: https://docs.espressif.com/projects/esp-idf/en/latest/esp32/hw-reference/esp32/user-guide-devkitm-1.html

[22] "ESP32-MINI-1 Datasheet", Aug. 2023. Accessed: Nov. 28, 2023. Available: https://www.espressif.com/sites/default/files/documentation/esp32-mini-1_datasheet_en.pdf

[23] "ESP32-C6 Series Datasheet", Jul. 2023. Accessed: Nov. 28, 2023. Available: https://www.espressif.com/sites/default/files/documentation/esp32-c6_datasheet_en.pdf

[24] "ESP32-C3 Series Datasheet", Aug. 2023. Accessed: Nov. 28, 2023. Available: https://www.espressif.com/sites/default/files/documentation/esp32-c3_datasheet_en.pdf

[25] "STM32CubeIDE - Integrated Development Environment for STM32". Accessed: Nov. 28, 2023. Available: https://www.st.com/en/development-tools/stm32cubeide.html

[26] "STM32 Peripheral Access Crates". Accessed: Nov. 28, 2023. Available: https://github.com/stm32-rs/stm32-rs

[27] "Bluetooth Low Energy and 802.15.4 Nucleo pack based on STM32WB Series microcontrollers", Sep. 2018. Accessed: Nov. 28, 2023. Available: https://www.st.com/resource/en/user_manual/um2435-bluetooth-low-energy-and-802154-nucleo-pack-based-on-stm32wb-series-microcontrollers-stmicroelectronics.pdf

[28] "STM32WB55xx Datasheet", Aug. 2023. Accessed: Nov. 28, 2023. Available: https://www.st.com/resource/en/datasheet/stm32wb55rg.pdf

[29] "Arduino Nano 33 BLE Product Reference Manual", Aug. 2022. Accessed: Nov. 28, 2023. Available: https://docs.arduino.cc/resources/datasheets/ABX00030-datasheet.pdf

[30] "nRF52840 Product Specifications", Feb. 2019. Accessed: Nov. 28, 2023. Available: https://infocenter.nordicsemi.com/pdf/nRF52840_PS_v1.1.pdf

[31] "Getting Started with Seeed Studio XIAO nRF52840 (Sense)". Accessed: Nov. 28, 2023. Available: https://wiki.seeedstudio.com/XIAO_BLE/

[32] "nRF7002 DK Hardware", May 2023. Accessed: Nov. 28, 2023. Available: https://infocenter.nordicsemi.com/pdf/nRF7002_DK_User_Guide_v1.0.0.pdf

[33] "nRF7002 Product Specifications", Feb. 2023. Accessed: Nov. 28, 2023. Available: https://infocenter.nordicsemi.com/pdf/nRF7002_PS_v1.0.pdf

[34] "nRF5340 DK Hardware", Jan. 2022. Accessed: Nov. 28, 2023. Available: https://infocenter.nordicsemi.com/pdf/nRF5340_DK_User_Guide_v2.0.0.pdf

[35] "nRF5340 Product Specifications", Oct. 2022. Accessed: Nov. 28, 2023. Available: https://infocenter.nordicsemi.com/pdf/nRF5340_PS_v1.3.pdf

[36] "Code Composer Studio integrated development environment (IDE)". Accessed: Nov. 28, 2023. Available: https://www.ti.com/tool/CCSTUDIO

[37] "SimpleLink low power software development kits (SDKs)". Accessed: Nov. 28, 2023. Available: https://www.ti.com/tool/SIMPLELINK-LOWPOWER-SDK

[38] "CC1352P7 SimpleLink High-Performance Multi-Band Wireless MCU With Integrated Power Amplifier", Nov. 2021. Accessed: Nov. 28, 2023. Available: https://www.ti.com/lit/pdf/SWRS251

[39] "MSP430FR604x, MSP430FR504x 16-MHz MCU up to 64KB FRAM, 12-Bit High-Speed 8-MSPS Sigma-Delta ADC, and Integrated Sensor AFE", Sep. 2021. Accessed: Nov. 28, 2023. Available: https://www.ti.com/lit/ds/slasef5b/slasef5b.pdf

[40] "TI E2E support forum for LP-CC1352P7". Accessed: Nov. 28, 2023. Available: https://e2e.ti.com/search?category=forum&q=LP-CC1352P7

[41] D. Blaauw *et al.*, "IoT design space challenges: Circuits and systems", *2014 Symposium on VLSI Technology (VLSI-Technology): Digest of Technical Papers*, Jun. 2014, doi: 10.1109/VLSIT.2014.6894411.

[42] E. A. Evangelakos, D. Kandris, D. Rountos, G. Tselikis, and E. Anastasiadis, "Energy Sustainability in Wireless Sensor Networks: An Analytical Survey", *Journal of Low Power Electronics and Applications*, vol. 12, no. 4, p. 65, Dec. 2022, doi: 10.3390/jlpea12040065.

[43] J. Singh, R. Kaur, and D. Singh, "Energy harvesting in wireless sensor networks: A taxonomic survey", *International Journal of Energy Research*, vol. 45, no. 1, pp. 118–140, Jan. 2021, doi: 10.1002/er.5816.

[44] A. Zahid Kausar, A. W. Reza, M. U. Saleh, and H. Ramiah, "Energizing wireless sensor networks by energy harvesting systems: Scopes, challenges and approaches", *Renewable and Sustainable Energy Reviews*, vol. 38, pp. 973–989, Oct. 2014, doi: 10.1016/j.rser.2014.07.035.

[45] M. Lv and E. Xu, "Deep Learning on Energy Harvesting IoT Devices: Survey and Future Challenges", *IEEE Access*, vol. 10, pp. 124999–125014, 2022, doi: 10.1109/ACCESS.2022.3225092.

[46] "WSN Solar Kit - Seeed Studio". Accessed: Nov. 28, 2023. Available: https://wiki.seeedstudio.com/Wireless_Sensor_Node-Solar_Kit/

[47] M. Gasulla and M. Carandell, "Power Gain from Energy Harvesting Sources at High MPPT Sampling Rates", *Sensors*, vol. 23, no. 9, p. 4388, Apr. 2023, doi: 10.3390/s23094388.

[48] A. Industries, "Adafruit Universal Solar Charger". Accessed: Nov. 28, 2023. Available: https://www.adafruit.com/product/4755

[49] "AEM10941 Evaluation Board User Guide", Sep. 2023. Accessed: Nov. 28, 2023. Available: https://e-peas.com/wp-content/uploads/2023/09/UG-AEM10941_QFN28_EVK2.5-v1.0.pdf

[50] Y. B. Zikria, S. W. Kim, O. Hahm, M. K. Afzal, and M. Y. Aalsalem, "Internet of Things (IoT) Operating Systems Management: Opportunities, Challenges, and Solution", *Sensors*, vol. 19, no. 8, p. 1793, Jan. 2019, doi: 10.3390/s19081793.

[51] "MyNewt open-source repository". Accessed: Nov. 28, 2023. Available: https://github.com/apache/mynewt-core

[52] "Nuttx open-source repository". Accessed: Nov. 28, 2023. Available: https://github.com/apache/nuttx

[53] "FreeRTOS open-source repository". Accessed: Nov. 29, 2023. Available: https://github.com/FreeRTOS/FreeRTOS-Kernel

[54] "Zephyr RTOS open-source repository". Accessed: Nov. 28, 2023. Available: https://github.com/zephyrproject-rtos/zephyr

[55] E. Baccelli *et al.*, "RIOT: An Open Source Operating System for Low-End Embedded Devices in the IoT". Aug. 13, 2023. doi: 10.1109/JIOT.2018.2815038.

[56] B. Gautier, "Operating system dedicated to NVRAM-based low power embedded systems", 2021. Accessed: Nov. 28, 2023. Available: https://theses.hal.science/tel-03192646

[57] M. Karimi, Y. Wang, Y. Kim, Y. Lim, and H. Kim, "CARTOS: A Charging-Aware Real-Time Operating System for Intermittent Batteryless Devices". arXiv, Nov. 2023. doi: 10.48550/arXiv.2311.07227.

[58] S. Umesh and S. Mittal, "A survey of techniques for intermittent computing", *Journal of Systems Architecture*, vol. 112, p. 101859, Jan. 2021, doi: 10.1016/j.sysarc.2020.101859.

[59] S. Lee, B. Islam, Y. Luo, and S. Nirjon, "Intermittent Learning: On-Device Machine Learning on Intermittently Powered System", no. arXiv:1904.09644. arXiv, Dec. 15, 2019. doi: 10.48550/arXiv.1904.09644.

[60] A. Bakar, A. G. Ross, K. S. Yildirim, and J. Hester, "REHASH: A Flexible, Developer Focused, Heuristic Adaptation Platform for Intermittently Powered Computing", *Proceedings of the ACM on Interactive, Mobile, Wearable and Ubiquitous Technologies*, vol. 5, no. 3, pp. 1–42, Sep. 2021, doi: 10.1145/3478077.